\documentclass[12pt]{article}

\usepackage{amsmath}
\usepackage{graphicx}
\usepackage{hyperref}
\usepackage[FIGTOPCAP]{subfigure}

\begin{document}

\title{Small oscillations of a heavy symmetric top}

\author{        V. Tanr{\i}verdi \\
tanriverdivedat@googlemail.com \\
Address: Bahad{\i}n Kasabas{\i} 66710 Sorgun-Yozgat TURKEY
}

\date{}

\maketitle

\begin{abstract}
This study is on small oscillations of a heavy symmetric top. 
A different method than previous works is applied, and differently from previous works, the explicit formulas for the amplitudes for oscillations are given.
This method can be helpful in teaching.
\end{abstract}

\section{Introduction}

Studying the motion of a heavy symmetric top by considering small oscillations is not a new topic, it dates back to $19^{th}$ century.
Different scientists had studied this topic from different perspectives.
Routh studied small oscillations of a heavy symmetric top by considering regular precession and oscillations of both inclination angle and precession angular velocity \cite{Routh}.
Crabtree also used a similar method and obtained the same result \cite{Crabtree}.
Lamb also considered similar arguments to Routh and obtained an equivalent result \cite{Lamb}.
Scarborough considered the motion with cusps and studied small oscillations \cite{Scarborough}.
There are also more recent works considering small oscillations of a heavy symmetric top.
Goldstein et al., like Scarborough, consider tops making the motion with cusps and look for the fast top limit \cite{Goldstein}.
Arnold also studies small oscillations of a heavy symmetric top considering motion with cusps but from a different aspect \cite{Arnold}.
Greiner considers the effective potential and obtains angular frequency for small oscillations \cite{Greiner}.

In general, a solution for the motion of a heavy symmetric top can be found either by using elliptic integrals or numerical integrations of angular accelerations \cite{Goldstein, Tanriverdi2019}.
These can be found abstract by students, especially when they first meet the concept of rigid body rotations.
And, as a result of this, they may find bizarre changes in angular velocities.
On the other hand, considering the motion in terms of small oscillations provides an approximate analytical solution, which can be used to show different aspects of the motion.
This can provide a nice path for teaching.

In this work, differently from previous works, we will use angular acceleration of the inclination angle to obtain equations for small oscillations.
This derivation is a bit easier to follow and it can increase the understandings of students.
Besides this, we will calculate amplitudes of oscillations, and 
we will also find changes in precession and spin angles and angular velocities in terms of small oscillations.
Scarborough studied motion with cups in a similar way and used this to plot the shapes for locus \cite{Scarborough}, and we will also plot them in a similar way.
We will consider two cases: The first one is easier and can be used while teaching to undergraduate students, and the second one can be used for upper-level undergraduate and graduate students.

In section \ref{one}, we will give basic equations for a heavy symmetric top which will be used for small oscillations.
In section \ref{two}, we will obtain equations for small oscillations.
In section \ref{three}, we will conclude.
In the appendix, we will give other relations for angular frequency.

\section{Heavy symmetric top}
\label{one}

For a heavy symmetric top, Lagrangian can be written as \cite{MarionThornton}
\begin{equation}
        L=\frac{I_x}{2}(\dot \theta ^2 + \dot \phi ^2 \sin^2 \theta)+\frac{I_z}{2}(\dot \psi+\dot \phi \cos \theta)^2-M g l \cos \theta, 
        \label{lagrngn}
\end{equation}
where $\theta$, $\phi$ and $\psi$ are Euler angles, and the time derivatives of them correspond angular velocities, $M$ is the mass, $l$ is the distance between the center of mass and the fixed point of the top, $I_x$ and $I_z$ are moments of inertia, and $g$ is the gravitational acceleration.

By using Euler-Lagrange equations, the angular acceleration $\ddot \theta$ can be obtained as \cite{Tanriverdi2019}
\begin{equation}
        \ddot \theta=\frac{\sin \theta}{I_x} \left[ Mgl+I_x \dot \phi^2 \cos \theta-I_z\dot \phi^2 \cos \theta-I_z \dot \phi \dot \psi \right], \label{ddottheta}
\end{equation}
and two conserved angular momenta can be obtained as
\begin{eqnarray}
        L_{z}&=&I_z(\dot \psi+\dot \phi \cos \theta), \nonumber \\
        L_{z'}&=&I_x \dot\phi \sin^2 \theta +I_z (\dot \psi+\dot \phi \cos \theta)\cos \theta .
        \label{angmom}
\end{eqnarray}
From these, one can define two constants: $a=L_z/I_x$ and $b=L_{z'}/I_x$ \cite{Goldstein}.
By using these constants, one can obtain $\dot \phi$ and $\dot \psi$ as
\begin{eqnarray}
	\dot \phi(\theta)&=&\frac{b - a \cos \theta}{\sin^2 \theta}, \label{phidot} \\
	\dot \psi(\theta)&=&\frac{I_x}{I_z}a-\frac{b - a \cos \theta}{\sin^2 \theta}\cos\theta. \label{psidot}
\end{eqnarray}
These two equations show that finding the change of $\theta$ means finding changes of $\dot \phi$ and $\dot \psi$.

One can define a constant from energy as
\begin{equation}
        E'=\frac{I_x}{2}\dot \theta ^2 +\frac{I_x}{2} \frac{(b-a \cos \theta)^2}{\sin^2 \theta}+M g l \cos \theta, \label{eprime}
\end{equation}
and by using this constant, one can define an effective potential \cite{Goldstein}
\begin{equation}
        U_{eff}(\theta)= \frac{I_x}{2}\frac{(b-a \cos \theta)^2}{\sin^2 \theta}+Mgl \cos \theta.
        \label{ueff}
\end{equation}
This effective potential goes to infinity at the domain boundaries of $\theta$ ($0$ and $\pi$), and it has a minimum between these when $|b|\ne|a|$.

\section{Small oscillations}
\label{two}

For a physical motion, $E'$ is either equal to the minimum of the effective potential or greater than that.
When $E'$ is slightly greater than $U_{eff_{min}}$, changes in $\theta$ become small and one can use small oscillation approximation.

We will take $I_x=I_y=22.8 \times 10^{-5}\, kg\, m^2$, $I_z=5.72 \times 10^{-5}\, kg\, m^2$ and $Mgl=0.068 J$ for the below examples.
We will compare the results with the ones obtained from numerical integrations of angular accelerations, the details of which can be found in previous work \cite{Tanriverdi2019}.

\subsection{Small oscillations for motion with cusps}

If one chooses initial conditions of a heavy symmetric top as $\dot \theta_0=0$ and $\dot \phi_0=0$ ($\theta_0 \ne 0$ and $\pi$, and $\dot \psi_0 \ne 0$), then $b$ becomes equal to $a\cos \theta_0$ and $E'=Mglb/a$.
In these conditions, motion with cusps is observed.

One can rewrite equation \eqref{ddottheta} in terms of $a$ and $b$ as 
\begin{equation}
	\ddot \theta=\frac{Mgl }{I_x} \sin \theta+\frac{(b-a \cos \theta)^2 \cos \theta}{\sin^3 \theta}-\frac{a(b-a \cos \theta)}{\sin \theta}. 
	\label{ddottheta2}
\end{equation}
By considering small oscillations and using $\theta=\theta_0+\delta$, one can approximate $\sin \theta\approx \sin \theta_0+\delta \cos \theta_0$ and $\cos \theta \approx \cos \theta_0-\delta \sin \theta_0$.
For small oscillations, one can expand the denominators of equation \ref{ddottheta2} and ignore higher-order terms in $\delta$. 
One can write $b=a \cos \theta_0$ for motions with cusps, and then one can obtain
\begin{equation}
	\ddot \delta + \left( a^2-\frac{Mgl}{I_x} \cos \theta_0 \right) \delta \approx \frac{Mgl}{I_x}\sin \theta_0.
\end{equation}
Here, one can define two constants of motion, which can be obtained from initial conditions as: $w_1^2=a^2-(Mgl/I_x) \cos \theta_0$ and $C_1=(Mgl/I_x) \sin \theta_0$.
The first one corresponds to angular frequency.
The left-hand side of this equation is the same as the harmonic oscillator and by using this property, one can write a solution as $\delta(t)=A \cos (w_1 t) + B \sin (w_1 t)+C_1/w_1^2$.
If initially $\dot \delta(t=0)=0$ and $\delta(t=0)=0$ or $\dot \theta(t=0)=0$ and $\theta(t=0)=\theta_0$, then one can get the amplitude of the oscilation as $A=-C_1/w_1^2$ and $B=0$, and the solution becomes 
\begin{equation}
	\delta(t)\approx \frac{C_1}{w_1^2}(1-\cos (w_1 t)).	
\end{equation}
Then, the solution for $\theta$ and angular velocities can be obtained as
\begin{eqnarray}
	\theta(t) &\approx& \theta_0 +\frac{C_1}{w_1^2}(1-\cos (w_1 t)),  \label{thetasc}\\
	\dot \theta(t) &\approx& \frac{C_1}{w_1} \sin (w_1 t), \\
	\dot \phi(t)&\approx& \frac{a}{\sin \theta_0}\frac{C_1}{w_1^2}(1-\cos (w_1 t)), \\
	\dot \psi(t)&\approx& \dot \psi_0-\frac{a \cos \theta_0}{\sin \theta_0}\frac{C_1}{w_1^2}(1-\cos (w_1 t)).
\end{eqnarray}
While obtaining solutions for $\dot \phi(t)$ and $\dot \psi (t)$, we have used equations \eqref{phidot} and \eqref{psidot} and applied small oscillations procedure.
One can also integrate $\dot \phi(t)$ and $\dot \psi(t)$ with respect to time and by considering $\phi(t=0)=0$ and $\psi(t=0)=0$, and find 
\begin{eqnarray}
	\phi(t)&\approx& \frac{a}{\sin \theta_0}\frac{C_1}{w_1^2} \left(t-\frac{\sin(w_1 t)}{w_1} \right), \label{phisc} \\
	\psi(t)&\approx& \dot \psi_0 t-\frac{a \cos\theta_0}{\sin\theta_0}\frac{C_1}{w_1^2} \left(t-\frac{\sin(w_1 t)}{w_1} \right).
\end{eqnarray}

When $|a| >> |Mgl|/I_x$, which is valid for very fast tops, one can approximate angular frequency as $a$ which we will designate as $w_2$.
This is an equivalent result to the ones given in classical mechanics books \cite{Lamb, Scarborough, Goldstein, Arnold}.  

\begin{figure}[h!]
\begin{center}
\subfigure[$\theta$]{
\includegraphics[width=4.4cm]{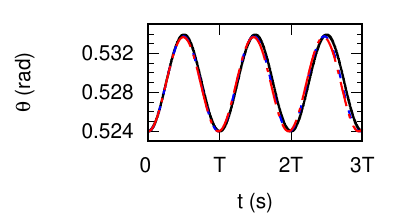}
}
\subfigure[$\dot \theta$]{
\includegraphics[width=4.4cm]{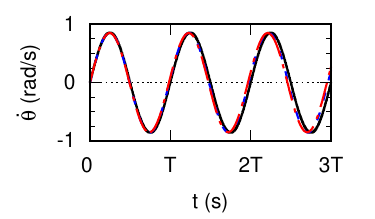}
}
\subfigure[$\dot \phi$]{
\includegraphics[width=4.4cm]{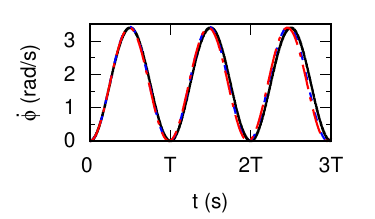}
}
\subfigure[$\dot \psi$]{
\includegraphics[width=4.4cm]{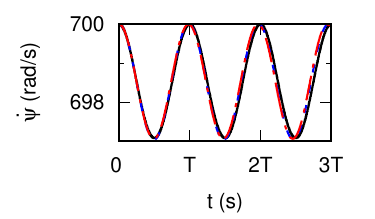}
}
\caption{
Time evolution of $\theta$ (a), $\dot \theta$ (b), $\dot \phi$ (c) and $\dot \psi$ (d) for motion with cusps.
        Continuous (black) curves show the results of numerical integration of angular accelerations, dotted-dashed (red) curves represent results for the small oscillations with $w_2$, dashed (blue) curves show results for the small oscillations with $w_1$.
        Initial values are $\theta_0=0.524\, rad$, $\dot \theta_0=0$, $\dot \phi_0=0 $ and $\dot \psi_0=700 \,rad\,s^{-1}$, and $T=0.0364\, s$.
        }
\label{fig:thetaphipsi_1}
\end{center}
\end{figure}

\begin{figure}[h!]
\begin{center}
\subfigure[]{
\includegraphics[width=4.6cm]{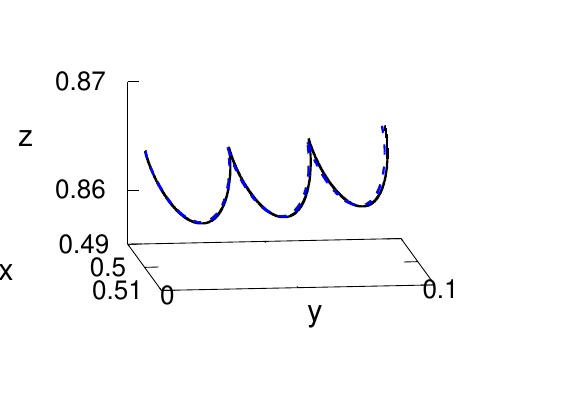}
}
\subfigure[]{
\includegraphics[width=4.6cm]{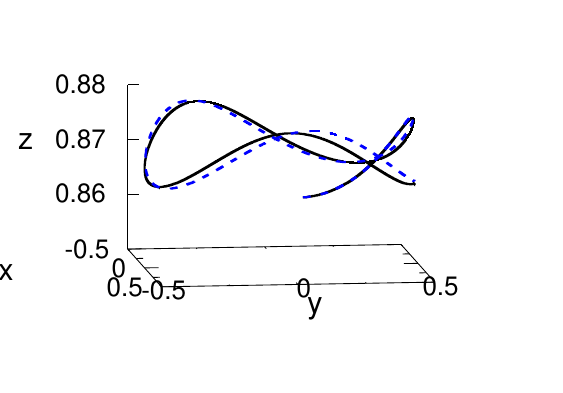}
}
\subfigure[]{
\includegraphics[width=4.6cm]{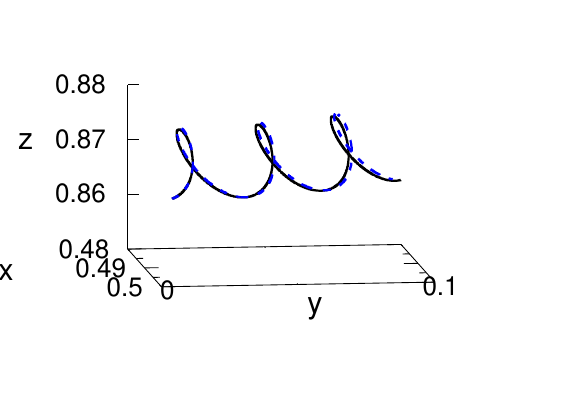}
}
\subfigure[]{
\includegraphics[width=4.6cm]{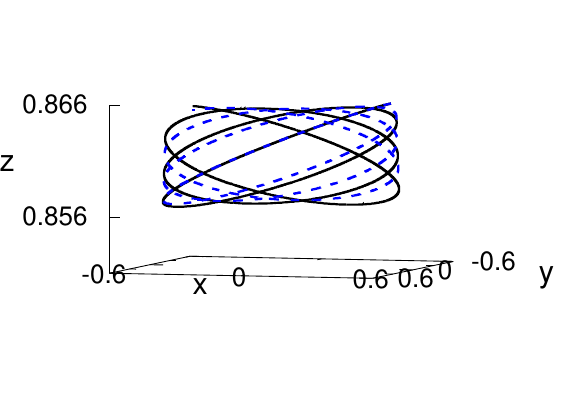}
}
\caption{
	Shapes for the locus. Continuous (black) curves show results of numerical integration of angular accelerations, and dashed (blue) curves show results of small oscillation approximation, i.e. obtained by using equations \eqref{thetasc} and \eqref{phisc} or \eqref{so1} and \eqref{so5}.
	(a) Motion with cusps, initial values are given in figure \ref{fig:thetaphipsi_1}. 
	(b) Wavy precession, initial values are given in figure \ref{fig:thetaphipsi_2}. 
	(c) Looping motion, initial values are given in figure \ref{fig:thetaphipsi_3}.
	(c) Motion when $b>a$, initial values are given in figure \ref{fig:thetaphipsi_4}.
        }
\label{fig:gr_tt}
\end{center}
\end{figure}

Now, we will consider an example and compare results for $w_1$ and $w_2$ with the solution obtained from numerical integrations of angular accelerations.
In this example, the initial conditions are chosen as $\dot \phi_0=0$, $\dot \theta_0=0$, $\theta_0=0.524\, rad$ and $\dot \psi_0=700 rad\,s^{-1}$.
Results of numerical integrations and small oscillations for $w_1$ and $w_2$ can be seen in figure \ref{fig:thetaphipsi_1}.
The shapes for the locus for this case can be seen in figure \ref{fig:gr_tt}(a), which are obtained both from small oscillation approximation with $w_1$ and numerical integration of angular accelerations.

From figure \ref{fig:thetaphipsi_1}, it can be seen that the small oscillation approximation gives close results to the numerical solution obtained from the integration of angular accelerations.
One can say that considering small oscillations gives slightly different angular frequency with respect to the integration of angular accelerations, 
i.e. $w_1=175 rad\,s^{-1} > w=173 rad\,s^{-1}$ , and slightly different amplitudes which are expected results due to ignoring higher-order terms.
The percentage difference in angular frequency is about $1.3\%$, and the percentage difference in the amplitude of oscillations of $\theta$ is $1.6\%$.

From figure \ref{fig:thetaphipsi_1}, it can be seen that ignoring $(Mgl/I_x) \cos \theta_0$ causes small differences.
As a result of this ignoring, the angular frequency becomes equal to $a$, i.e. $w_2=176 rad\,s^{-1}$, and the percentage difference is slightly greater, i.e. $1.7\%$.
The percentage difference for the amplitude of oscillations of $\theta$ is the same since we have used the same amplitude.

As $|\dot \psi_0|$ and dependently $|a|$ increase, the percentage differences become smaller.
And, there are two reasons for this:
One of them is that as $|a|$ increases the contribution from $Mgl \cos \theta$ becomes smaller which results in the increase of the resemblance of effective potential to the potential of the harmonic oscillator as long as $|b|$ is not close to $|a|$.
The other one is related to difference between $E'=Mglb/a$ and $U_{eff_{min}}$, which becomes smaller as $|a|$ increases \cite{Tanriverdi_ueff}.

\subsection{Small oscillations in different types of motion}

Now, we will study the small oscillations of a heavy symmetric top in general, i.e. $b = a \cos \theta_0$ condition does not need to be satisfied.
By considering small oscillations similar to the previous case, one can write $\theta=\theta_0+\eta$.
Then, from equation \eqref{ddottheta2}, one can obtain
\begin{eqnarray}
        \ddot \eta &\approx& \frac{Mgl}{I_x} \sin \theta_0+\frac{(a^2+b^2) \cos \theta_0-a b (1+\cos^2 \theta_0)}{\sin^3 \theta_0}   \label{approx1} \\
        & &+\eta\left[ \frac{Mgl}{I_x} \cos \theta_0+\frac{a b \cos \theta_0 (5+\cos^2 \theta_0)-(a^2+b^2) (1+2 \cos^2 \theta_0)}{\sin^4 \theta_0}\right]. \nonumber 
\end{eqnarray}
Then, one can define
\small
\begin{eqnarray}
	w_g^2&=&-\frac{Mgl}{I_x} \cos \theta_0-\frac{a b \cos \theta_0 (5+\cos^2 \theta_0)-(a^2+b^2) (1+2 \cos^2 \theta_0)}{\sin^4 \theta_0}, \\
	C_g&=&\frac{Mgl}{I_x} \sin \theta_0+\frac{(a^2+b^2) \cos \theta_0-a b (1+\cos^2 \theta_0)}{\sin^3 \theta_0}.
\end{eqnarray}
\normalsize
Then, one can write $\ddot \eta +w_g^2 \eta \approx C_g$ whose solution can be written as $\eta(t)=A \cos (w_g t)+B \sin(w_g t)+C_g/w_g^2$.
Similar to previous case, one can consider the case $\theta(t=0)=\theta_0$ and $\dot \theta (t=0)=0$ and obtain for small oscillations
\begin{eqnarray}
	\theta(t) &\approx& \theta_0+\frac{C_g}{w_g^2}(1- \cos(w_g t)), \label{so1}\\
	\dot \theta(t) &\approx& \frac{C_g}{w_g} \sin (w_g t), \label{so2}\\
	\dot \phi(t) &\approx& \dot \phi_0 + \frac{a-2 \dot \phi_0 \cos \theta_0}{\sin \theta_0}\frac{C_g}{w_g^2}(1- \cos(w_g t)), \label{so3}\\
	\dot \psi(t) &\approx& \dot \psi_0 + \frac{2 \dot \phi_0 -b}{\sin \theta_0}\frac{C_g}{w_g^2}(1- \cos(w_g t)). \label{so4}
\end{eqnarray}
Both angular frequency $w_g$ and $C_g$ can be obtained from initial values so do amplitudes for all oscillations.
Similar to previous case, one can integrate $\dot \phi(t)$ and $\dot \psi(t)$ with respect to time and find 
\begin{eqnarray}
	\phi(t)&\approx& \dot \phi_0 t+\frac{a-2 \dot \phi_0 \cos \theta_0}{\sin \theta_0 }\frac{C_g}{w_g^2}\left(t-\frac{\sin(w_g t)}{w_g} \right), \label{so5} \\
	\psi(t)&\approx& \dot \psi_0 t+\frac{2 \dot \phi_0-b}{\sin\theta_0}\frac{C_g}{w_g^2}\left(t-\frac{\sin(w_1 t)}{w_1}\right), \label{so6}
\end{eqnarray}
where $\phi(t=0)=0$ and $\psi(t=0)=0$, again.

\begin{figure}[h!]
\begin{center}
\subfigure[$\theta$]{
\includegraphics[width=4.4cm]{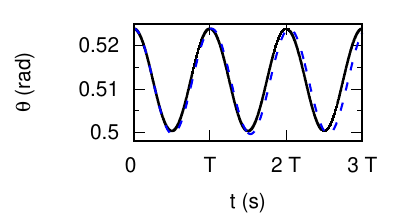}
}
\subfigure[$\dot \theta$]{
\includegraphics[width=4.4cm]{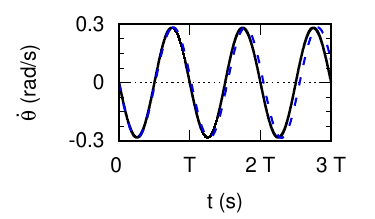}
}
\subfigure[$\dot \phi$]{
\includegraphics[width=4.4cm]{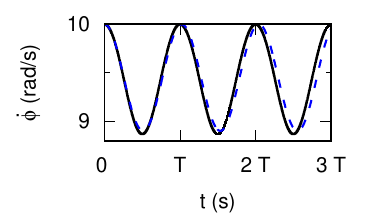}
}
\subfigure[$\dot \psi$]{
\includegraphics[width=4.4cm]{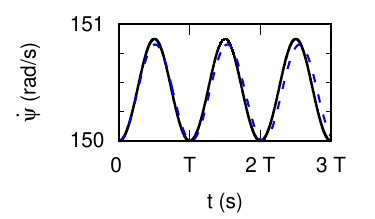}
}
\caption{
Time evolution of $\theta$ (a), $\dot \theta$ (b), $\dot \phi$ (c) and $\dot \psi$ (d) for wavy precession.
	Continuous (black) curves show the results of numerical integration of angular accelerations, dashed (blue) curves show results of equations \ref{so1}, \ref{so2}, \ref{so3} and \ref{so4}.
        Initial values are $\theta_0=0.524\, rad$, $\dot \theta_0=0$, $\dot \phi_0=10 \,rad\,s^{-1}$ and $\dot \psi_0=150 \,rad\,s^{-1}$, and $T=0.264\,s$.
        }
\label{fig:thetaphipsi_2}
\end{center}
\end{figure}

Now, we will consider three different examples: Wavy precession, looping motion and a motion when $b>a$.
Shapes for the locus, obtained from both numerical integration of angular accelerations and small oscillations for these examples, can be seen in figure \ref{fig:gr_tt}.

The first example is wavy precession which is the most possible motion for small oscillations when $|a|>|b|$,
since $U_{eff_{min}}<E'<Mglb/a$ in this case \cite{Tanriverdi_abdffrnt}.
Initial values of this example can be found in the explanations of figure \ref{fig:thetaphipsi_2}.
From that figure, it can be seen that results of equations \eqref{so1}, \eqref{so2}, \eqref{so3} and \eqref{so4} are close to the results of numerical integrations of angular accelerations.
For this example, $w_g=23.3 rad\,s^{-1}$ and the angular frequency obtained from numerical integration is $w=23.8 rad\,s^{-1}$, and the percentage difference is $2.1\%$, which is $0.2\%$ for the amplitude of $\theta$. 

\begin{figure}[h!]
\begin{center}
\subfigure[$\theta$]{
\includegraphics[width=4.4cm]{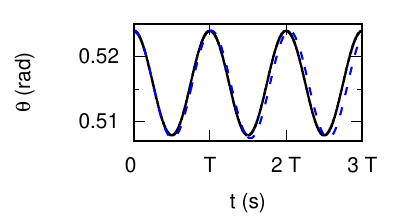}
}
\subfigure[$\dot \theta$]{
\includegraphics[width=4.4cm]{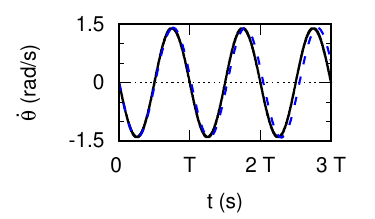}
}
\subfigure[$\dot \phi$]{
\includegraphics[width=4.4cm]{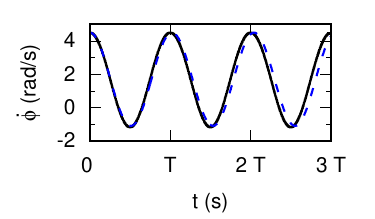}
}
\subfigure[$\dot \psi$]{
\includegraphics[width=4.4cm]{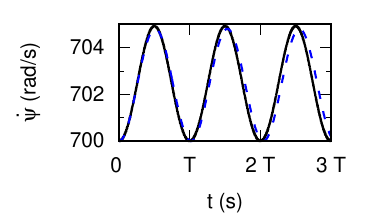}
}
\caption{
Time evolution of $\theta$ (a), $\dot \theta$ (b), $\dot \phi$ (c) and $\dot \psi$ (d) for looping motion.
	Legends are the same as figure \ref{fig:thetaphipsi_2}.
        Initial values are $\theta_0=0.524\, rad$, $\dot \theta_0=0$, $\dot \phi_0=4.5  \,rad\,s^{-1} $ and $\dot \psi_0=700 \,rad\,s^{-1}$, and $T=0.0362\, s$.
        }
\label{fig:thetaphipsi_3}
\end{center}
\end{figure}

The second example of this part is looping motion which is seen when $E'>Mglb/a$ ($|a|>|b|$) \cite{Tanriverdi_abdffrnt}.
Then, to obtain an adequate example for looping motion with small oscillations, one may need to choose initial conditions resulting in greater $|a|$ values.
This is the reason for choosing a greater $\dot \psi_0$ value, which can be found in the explanations of figure \ref{fig:thetaphipsi_3} together with other initial values.
From that figure, it can be seen that one can obtain an approximate solution for looping motion by using small oscillation approximation.
For this example, $w_g=170 rad\,s^{-1}$ and $w=174 rad\,s^{-1}$, and the percentage difference is $2\%$, which is $1.9\%$ for the amplitude of $\theta$.

\begin{figure}[h!]
\begin{center}
\subfigure[$\theta$]{
\includegraphics[width=4.4cm]{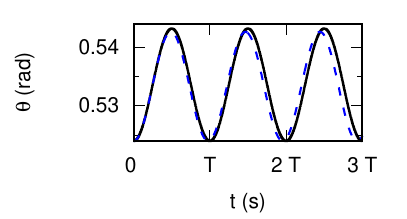}
}
\subfigure[$\dot \theta$]{
\includegraphics[width=4.4cm]{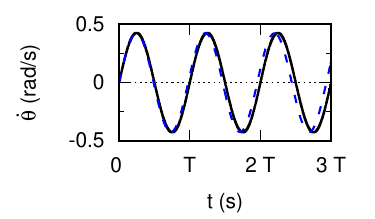}
}
\subfigure[$\dot \phi$]{
\includegraphics[width=4.4cm]{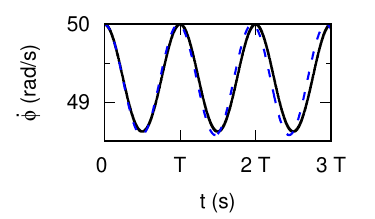}
}
\subfigure[$\dot \psi$]{
\includegraphics[width=4.4cm]{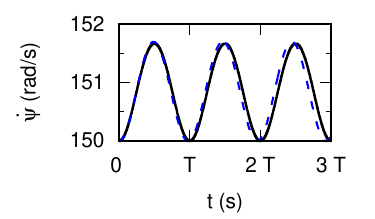}
}
\caption{
Time evolution of $\theta$ (a), $\dot \theta$ (b), $\dot \phi$ (c) and $\dot \psi$ (d) for the motion when $b>a$.
	Legends are the same as figure \ref{fig:thetaphipsi_2}.
        Initial values are $\theta_0=0.524\, rad$, $\dot \theta_0=0$, $\dot \phi_0=50 \,rad\,s^{-1}$ and $\dot \psi_0=150 \,rad\,s^{-1}$, and $T=0.142\, s$.
        }
\label{fig:thetaphipsi_4}
\end{center}
\end{figure}

The third example of this part is a motion when $b>a$.
The shapes for the locus for these types of motions are always similar \cite{Tanriverdi_abdffrnt}.
Results of numerical integrations of angular accelerations and small oscillation approximation can be seen in figure \ref{fig:thetaphipsi_4} where one can also find initial values.
One can see from that figure that the results of small oscillation approximation are close to numerical integrations's.
For this example, $w_g=45.2 rad\,s^{-1}$ and $w=42.3 rad\,s^{-1}$, and the percentage difference is $7\%$, which is $3.1\%$ for the amplitude of $\theta$.

\section{Conclusion}
\label{three}

In this work, we studied small oscillations of a heavy symmetric top.
We followed a different way from the previous works and obtained an equivalent result for angular frequency to that of the previous works, which can be found in the appendix.
In some cases, other forms of angular frequency can be more convenient and can be used.
In this work, we used the form given above to keep the derivation simple.
We should note that solutions are obtained when $\dot \theta_0=0$, and for other cases, the solutions should be modified.

We have obtained equations describing the changes in angles and angular velocities for small oscillations.
We have seen that angular velocities change as $\theta$ changes.
These changes in angular velocities occur due to the conservation of angular momenta and energy as they should be.
These conservation laws can not be explicitly seen from these equations.
But, we know that these equations are obtained from the equations of motion by using Lagrangian, which is time-independent and energy is conserved,
and, in fact, two of the equations of motion are noting but conservation of angular momenta for this case.
On the other hand, due to considering small oscillations and ignoring higher-order terms, these equations approximately satisfy conservation laws and they are valid up to some point,
for total concordance, naturally, one needs to consider full equations.

We have considered different types of motion and chosen moderate examples, for which percentage differences are less than $10\%$.
However, one can obtain examples with less than $1\%$ percentage differences when the top is fastly spinning, i.e. $|a|$ is much greater than used values.
On the other hand, as $|a|$ decreases or the difference between $E'$ and $U_{eff_{min}}$ increases the difference becomes bigger, and for larger values, small oscillation approximation can not be used as expected.
Shapes for the locus can also be approximately obtained from small oscillations by the formulas received for $\theta$ and $\phi$, i.e. equations \eqref{thetasc} and \eqref{phisc} or \eqref{so1} and \eqref{so5}, which can be helpful in teaching.
The studied examples show that both the angular frequency and amplitudes are convenient for small oscillations of a heavy symmetric top, and
studying different types of motion as examples shows that they can be used in different types of motion.
We should note that this small oscillation approximation also works for the negative $Mgl$ case.

\section{Appendix}

Routh approaches the small oscillations from a different perspective and takes into account small oscillations both in $\theta$ and $\dot \phi$ by considering small oscillations around the regular precession \cite{Routh} occurring at the minimum of the effective potential.
By assuming the form of the solution as $\eta_R=F \sin(w_R t+f)$, Routh finds the square of the angular frequency for small oscillations of $\theta$ which can be written in terms of parameters of this work as
\begin{equation}
	w_R^2=\dot \phi_{rp}^2-\frac{2mgl}{I_x} \cos \theta_{rp}+\left( \frac{mgl}{I_x \dot \phi_{rp}}\right)^2, \label{Rw}
\end{equation}
where $\theta_{rp}$ is the angle giving regular precession which is the angle making the effective potential minimum, and $\dot \phi_{rp}$ is the precession angular velocity corresponding to that regular precession.
Crabtree obtains the same result with Routh \cite{Crabtree}.

Lamb follows a similar path to Routh's approach and finds angular frequency which can be written as \cite{Lamb}
\begin{equation}
	w_L^2=\dot \phi_{rp}^2 \sin^2 \theta_{rp}+(a-2 \dot \phi_{rp} \cos \theta_{rp})^2. \label{Lw}
\end{equation}

Greiner obtains angular frequency from the second derivative of effective potential with respect to $\theta$, which can be written as \cite{Greiner}
\begin{equation}
	w_G^2=\frac{I_x a b-Mgl(4-3 \sin^2 \theta_{rp})}{I_x \cos \theta_{rp}}. \label{Gw}
\end{equation}

One can show that these three results are the same as $w_g^2$ by using equations \eqref{ddottheta}, \eqref{phidot} and \eqref{psidot} by considering $\ddot \theta=0$ for regular precession.

\end{document}